\documentclass[a4paper]{jpconf}
\usepackage{graphicx}
\begin{document}
\title{Total interference between nuclear and magnetovibrational one-phonon scattering cross sections}

\author{S. Raymond$^1$, N. Biniskos$^{1,2}$, K. Schmalzl$^2$, J. Persson$^{3}$ and T. Br\"uckel$^3$}

\address{$^1$ Univ. Grenoble Alpes, CEA, INAC-MEM, 38000 Grenoble, France}
\address{$^2$ Forschungszentrum J\"ulich GmbH, JCNS at ILL, 38000 Grenoble, France}
\address{$^3$ Forschungszentrum J\"ulich GmbH, JCNS-2 and PGI-4, JARA-FIT, 52425 J\"ulich, Germany}

\ead{raymond@ill.fr, n.biniskos@fz-juelich.de}

\today

\begin{abstract}
A full phonon intensity cancellation is reported in a longitudinal polarized inelastic neutron scattering experiment performed on the magnetocaloric compound MnFe$_{4}$Si$_{3}$, a ferromagnet with $T_{Curie}$ $\approx$ 305 K.
The TA[100] phonon polarized along the $c$-axis measured from the Brillouin zone center $\textbf{G}$=(0, 0, 2) is observed only in one ($\sigma_{z}^{++}$) of the two non-spin-flip polarization channels and is absent in the other one ($\sigma_{z}^{--}$) at low temperatures. 
This effect disappears at higher temperatures, in the vicinity of $T_{Curie}$, where the phonon is measured in both channels with nonetheless marked different intensities.
The effect is understood as originating from nuclear-magnetic interference between the nuclear one-phonon and the magnetovibrational one-phonon scattering cross-sections.
The total cancellation reported is accidental, i.e. does not correspond to a systematic effect, as established by measurements in different Brillouin zones.
\end{abstract}

\section{Introduction}
Interest for the magnetocaloric effect has been increasing in the last decades in view of applications for room temperature refrigeration \cite{Pecha}.
Neutron scattering studies can bring important microscopic information on the spin and lattice dynamics of these systems and reveal key ingredients at play in their magneto-thermodynamics properties \cite{Nils,Biniskos1,Biniskos2}.
In  such context, experiments were performed on the ferromagnetic compound MnFe$_{4}$Si$_{3}$ and the importance of short range magnetic correlations as well as their suppression by a modest magnetic field was highlighted \cite{Biniskos1}.
Besides these results on the spin dynamics, unusual features of the phonon intensities were found in longitudinal polarized neutron scattering experiments under magnetic field. 
The aim of the present paper is to describe these effects and to show that they can be explained by the neutron scattering cross-sections. 

\section{Experimental results}
The inelastic neutron scattering (INS) experiments were performed on the cold neutron three-axis spectrometer IN12 \cite{Schmalzl}.
The spectrometer was set in W configuration with a fully focusing setup. 
The incident neutron beam spin state was prepared with a transmission polarizing cavity located after the velocity selector and the Heusler analyzer was set at fixed $k_f$=2 {\AA}$^{-1}$.
The sample is the same single crystal as the one used in Ref.\cite{Biniskos1}. It was placed in a vertical field magnet, the field being along the $b$-axis of the hexagonal structure and the horizontal scattering plane being thus defined by $a^*$ and $c^*$.
A magnetic field of 1 T was applied in the paramagnetic state and the sample was field-cooled in order to get a single domain ferromagnetic sample. Hence, the polarization at the sample position is kept along the magnetic field which defines the $z$-axis.
Two Mezei spin flippers were used before and after the sample in order to measure the four possible polarized neutron cross-sections $\sigma^{\alpha \beta}_{z}$=\{$\sigma^{++}_{z}$ , $\sigma^{+-}_{z}$ , $\sigma^{--}_{z}$ , $\sigma^{-+}_{z}$ \}.  On IN12, "+" corresponds to "flipper on" since both the transmitting polarizer and the Heusler analyzer transport "-" state.  With such a setup, the flipping ratio measured on a Bragg reflection of a graphite sample was on average of 18 $\pm$ 3, the most important source of uncertainty arising from the setting of subsequent experiments. In the present paper, we do not apply polarization corrections to the data. This is justified by the fact that the leakage between the polarization channels due to the finite polarization is hardly observable with respect to the limited statistics of the inelastic signal (see below) and that the paper focuses on large effects.
\begin{figure}[h]
\centering
\vspace{-1cm}
\includegraphics[width=22cm]{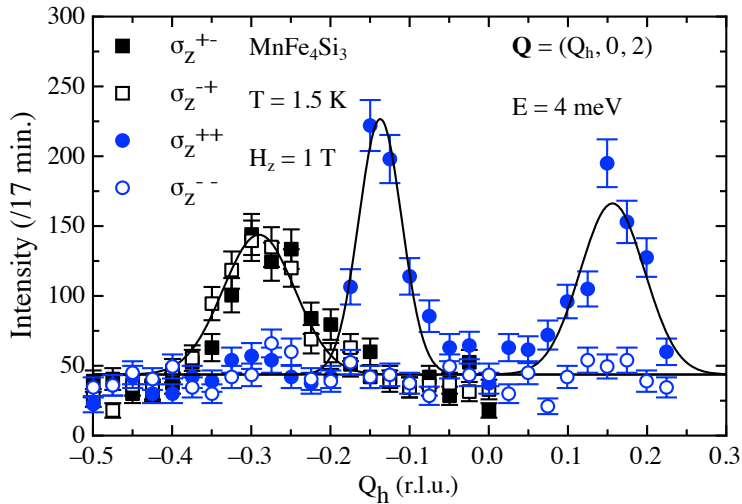}
\vspace{-9cm}
\caption{Inelastic neutron scattering spectra obtained in the four polarization channels $\sigma_z^{\alpha \beta}$ for $\textbf{Q}$=($Q_h$, 0, 2) for an energy transfer of 4 meV at $T$=1.5 K. Solid lines are Gaussian fits.}
\end{figure}

Figure 1 shows constant energy scans performed along the direction $\textbf{Q}$=($Q_h$, 0, 2) for an energy transfer $E$= 4 meV at $T$= 1.5 K for the four polarization channels $\sigma_z^{\alpha \beta}$.
For the spin-flip scattering, the peak at $Q_h$ $\approx$ -0.27 is observed with the same intensity and lineshape in the $\sigma^{+-}_{z}$ and $\sigma^{-+}_{z}$channels. It is ascribed to a spin-wave mode from the facts that (i) the direction of fluctuations is  found to be perpendicular to the ordered magnetic moments and (ii) its peak position is in agreement with the established magnon dispersion of MnFe$_{4}$Si$_{3}$ \cite{Biniskos1}.
The leakage of the magnon scattering from $\sigma^{\alpha \beta}_{z}$ ($\alpha \neq \beta$) to  $\sigma^{\alpha \alpha}_{z}$ due to finite polarization is barely seen above the background level for both channels.
In the non-spin-flip channels, a phonon mode is measured at $Q_h$ $\approx$ $\pm$ 0.15 in agreement with inelastic X-ray scattering data \cite{Biniskos3}.
Surprisingly this phonon mode is only seen in the $\sigma^{++}_{z}$ channel and has zero intensity in the $\sigma^{--}_{z}$ channel independent of the focusing ($Q_h$ $\approx$ - 0.15) or defocusing ($Q_h$ $\approx$ + 0.15) side of the measurement.

\begin{figure}[h]
%\centering
\vspace{-1cm}
\includegraphics[width=18cm]{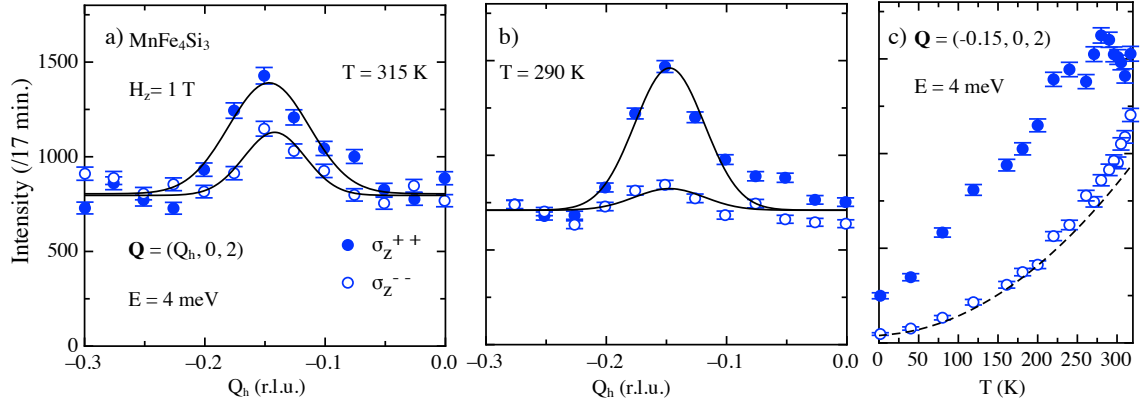}
\vspace{-8cm}
\caption{Inelastic neutron scattering spectra obtained in the $\sigma^{++}_{z}$ and $\sigma^{--}_{z}$ polarization channels for $\textbf{Q}$=($Q_h$, 0, 2) for an energy transfer of 4 meV at a) $T$=315 K and b) $T$=290 K. Solid lines are Gaussian fits. c) Temperature dependence of the peak intensity at $\textbf{Q}$=(-0.15, 0, 2) for an energy transfer of 4 meV. The dashed line indicates the background extrapolated from measurements performed at $\textbf{Q}$=(-0.3, 0, 2).}
\end{figure}

To get insight into this behavior, the temperature dependence of the phonon cross-sections $\sigma^{++}_{z}$ and $\sigma^{--}_{z}$  was studied. 
Fig.2 shows similar INS spectra as those of Fig.1 at $T$=315 K (Fig.2a) and $T$= 290 K (Fig.2b), two temperatures above and below $T_{Curie}$. In contrast to the low temperature data, the phonon appears in the two polarization channels at 315 K ; nevertheless the two phonon intensities differ significantly. At 290 K, only a tiny signal is observed in the $\sigma^{--}_{z}$ channel. 
Within our statistics, the phonon peaks at the same position and has the same width at low and high temperatures.
Fig.2  c) shows the temperature dependence of the phonon peak intensity in both polarization channels, where the temperature dependence of the lattice parameter was measured and taken into account.
The dashed line indicates the temperature dependence of the background determined by extrapolating measured values at low and high temperatures at $\textbf{Q}$=(-0.3, 0, 2) and $E$=4 meV.
From this curve, one can conclude that the phonon intensity in the $\sigma^{--}_{z}$ channel reaches the background level at around 200 K upon cooling from the paramagnetic state.

\section{Data interpretation}
The general polarized neutron scattering cross-sections indicate that the asymmetry  between $\sigma^{++}_{z}$  and $\sigma^{--}_{z}$  comes from the real part of the Nuclear Magnetic Interference (NMI) term, $R_z$ :
\begin{eqnarray}
\sigma^{++}_{z}=NN+M_zM_z+R_z\\
\sigma^{--}_{z}=NN+M_zM_z-R_z
\end{eqnarray}
where $NN$ and $M_zM_z$ are convenient notations adapted from Ref.\cite{Regnault} for the purely nuclear and purely magnetic correlation functions.
A well-known use of the NMI lies in the polarized neutron diffraction experiments aiming to measure magnetic form factors and spin densities \cite{Roessli}. NMI counterparts for inelastic neutron scattering are scarce.

In the range of wave-vectors where the phonon is measured in our experiment, there are no spin-waves and no diffuse spin fluctuations. Therefore, the only possible mixed nuclear-magnetic correlation function involves the uniform static magnetization of the system.
In this case, the NMI is elastic in the magnetic system and inelastic with respect to atomic vibrations. 
This suggests that it originates from the magnetovibrational scattering\footnote{It is recalled that, under the the assumption that the the motion of an ion is uncorrelated with either its spin direction or  magnitude, the magnetic scattering is composed of four terms : elastic magnetic scattering, magnetovibrational scattering, inelastic magnetic scattering (without change in the phonon system) and scattering which is inelastic in both the spin and the phonon systems \cite{Lovesey,Squires}.}, the creation (or annihilation) of a phonon via the magnetic interaction. 

In a real experiment, both cross-sections, the nuclear and the magnetovibrational one, participate to the scattering with their own magnitude.
Moreover, when longitudinal neutron polarization is used with a component of the polarization perpendicular to the scattering vector, both scattering amplitudes interfere with each others.
The polarized neutron cross-section including both nuclear and magnetic scattering are given in Ref.\cite{Lovesey} Eq.(10.157) and Ref.\cite{Squires} Eq.(9.49). 
Considering only the coherent vibrational cross-sections (nuclear and magnetovibrational), i.e. neglecting the inelastic magnetic scattering (and all elastic scattering) and considering the experimental case of a non-Bravais lattice in a single domain ferromagnetic state with perfect polarization along $z$, this simplifies to :
\begin{eqnarray}
\left(\frac{d^2\sigma}{d\Omega dE'}\right)^{\pm \pm}  \propto \int_{-\infty}^{+\infty}dt e^{-i\omega t}\times\sum_{j,j'}\left(\bar b_j\bar b_{j'}+\frac{(\gamma r_0)^2}{4}g_jg_{j'}F_j(\textbf{Q})F_{j'}(\textbf{Q}) \langle S^z_j\rangle \langle S^z_{j'}\rangle \pm \frac{\gamma r_0}{2}\bar b_{j}g_{j'}F_{j'}(\textbf{Q})\langle S^z_{j'}\rangle \right)\times\nonumber\\
\left(I_{j,j'}(\textbf{Q},t)-I_{j,j'}(\textbf{Q},\infty)\right)
\end{eqnarray}
with 
\begin{equation}
I_{j,j'}(\textbf{Q},t)=\langle \mathrm{exp}(-i\textbf{Q}\cdot\textbf{R}_{j}(0))\mathrm{exp}(i\textbf{Q}\cdot\textbf{R}_{j'}(t))\rangle
\end{equation}
where $j$ labels the atoms at the position $\textbf{R}_j(t)$ with a coherent scattering length $\bar b_j$ and the subset of magnetic sites have a form factor $F_j(\textbf{Q})$, a Land\'e splitting factor $g_j$ and a mean value of the $z$ component of the spin $\langle S^z_j\rangle$, $\gamma$ is the neutron gyromagnetic ratio and $r_{0}$ is the classical radius of the electron.
One can recognize the structure of (1)-(2) in (3). The simplicity of the magnetovibrational NMI term, being linear in $N$ and $M_z$, is shared with the well-known elastic NMI. 
It is not a general feature, the structure of $R_z$ can be much more complex especially for inelastic magnetic scattering.
Performing the canonical phonon expansion of Eq.(4) for a non-Bravais lattice \cite{Schober}, the polarized neutron one-phonon vibrational scattering double differential cross-section can then be written as  :
\begin{equation}
\left(\frac{d^2\sigma}{d\Omega dE'}\right)^{\pm \pm}  \propto \sum_{\textbf{G}} \sum_{s} \bigg|\sum_{d}\left(\bar b_d\pm\frac{\gamma r_{0}}{2}g_dF_d(\textbf{Q})\langle S^z_d\rangle\right)e^{-W_{d}(\textbf{Q})}e^{i\textbf{Q}\cdot{\textbf{d}}}(\textbf{Q}\cdot\textbf{e}_{ds})\frac{1}{\sqrt{m_{d}}}\bigg|^2\times F_s(\textbf{Q}, \omega, T)
\end{equation}
with the spectral weight function for a phonon creation process : 
\begin{eqnarray}
F_s(\textbf{Q}, \omega, T)=\frac{1}{\omega_{s}}\langle n_s+1 \rangle\delta(\omega-\omega_{s}) \delta(\textbf{Q}-\textbf{q}-\textbf{G})
\end{eqnarray}
where $s$ labels the phonon mode, $d$ labels the atoms in the unit cell, their mass is $m_d$, $e^{-W_{d}(\textbf{Q})}$ is the Debye-Waller factor, $\mathbf{e}_{ds}$ is the mode eigenvector and $\omega_{s}$ its frequency and $\langle n_s+1 \rangle$ is the thermal population factor.
$\textbf{G}$ is a Brillouin zone center and the scattering vector is written $\textbf{Q}$=$\textbf{G}$+$\textbf{q}$.
Eqn. (5) corresponds to the usual phonon scattering cross-section where the coherent scattering length is replaced by $\bar b_d\pm\frac{\gamma r_{0}}{2}g_dF_d(\textbf{Q})\langle S^z_d \rangle$.
For a phonon mode investigated in a restricted portion of the reciprocal space (typically half of a Brillouin zone length), one can make the approximation $F_d(\textbf{Q})$ $\approx$ $F_d(\textbf{G})$ and the scattering length is then $\textbf{q}$ independent in this range of wave-vectors.
To be precise, the occurrence of non zero NMI is made possible, in the case  described here, by the fact that the external magnetic field breaks the time reversal symmetry and creates a single domain ferromagnetic sample. 
The exact same interference effect has been used in the past in order to measure the "dynamical magnetic form factor" at finite $\textbf{q}$ using phonon scattering \cite{Steinsvoll}. 
In the present paper, we formulate this largely overlooked effect in the more modern viewpoint of NMI and extend the formulation for a non-Bravais lattice (See Eq.(1) of Ref.\cite{Steinsvoll} for a Bravais lattice).
In our experiment on MnFe$_{4}$Si$_{3}$, this effect is the best candidate to explain the cancellation of the phonon intensity at low temperatures.

\begin{figure}[h]
\centering
\vspace{-1cm}
\includegraphics[width=20cm]{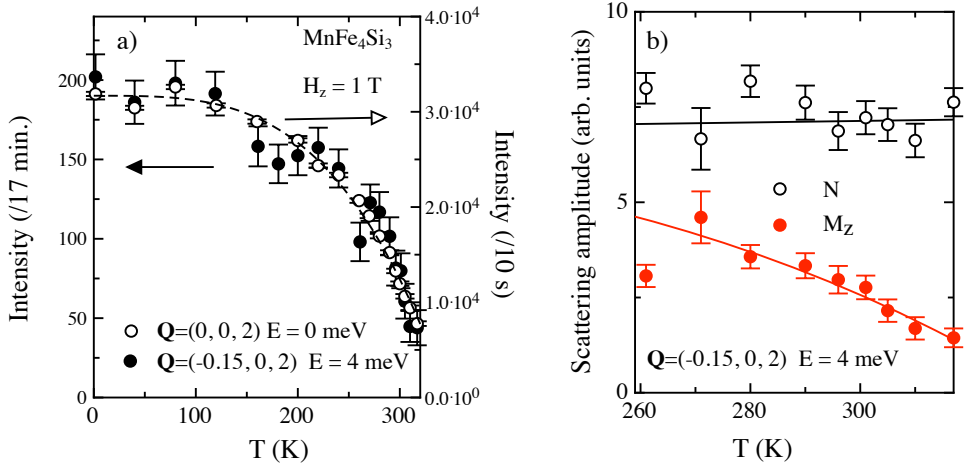}
\vspace{-9cm}
\caption{a) Temperature dependence of the interference term, 2$R_z$, measured on the phonon at $\bf{Q}$= (-0.15, 0, 2) at $E$ = 4 meV(full circles) and on the Bragg peak at $\bf{Q}$=(0, 0, 2) (open circles). b) Temperature dependence of the nuclear and magnetovibrational scattering amplitudes obtained from the phonon intensity at $\bf{Q}$= (-0.15, 0, 2) at $E$ = 4 meV. Lines are guides for the eyes.}
\end{figure}

In order to confirm this hypothesis and to describe the temperature dependence of the interference, Fig.3 shows $2R_z$ obtained by subtracting (1)-(2) using the data measured at $\textbf{Q}$=(-0.15, 0, 2) and $E$ = 4 meV and shown in Fig2.b.
In such a difference, the background, being equal in the $\sigma^{++}_z$ and $\sigma^{--}_z$ channels, cancels. The difference shown in Fig.3 is also corrected by the temperature population factor. 
$2R_z$ follows an order-parameter-like curve. 
Similarly, one can also obtain the elastic NMI term for the Bragg peak located at $\textbf{Q}$=(0, 0, 2) by subtracting the intensity measured in the $\sigma^{--}_z$ from the one measured in the $\sigma^{++}_z$ channel at this position for $E$=0 meV.
The corresponding data are also shown in the same figure. The two temperature dependences overlay within a single scaling factor, which confirms the static nature of the magnetic part of the interference term in the phonon scattering.
Indeed in the acoustic approximation, the dynamical structure factor is proportional to the static one  \cite{Xu}. Therefore their temperature dependence also matches when the phonon energy does not vary with temperature and when the thermal population factor is taken into account.
In general, the overlay shown in Fig.3a should not be completely quantitative since the Bragg peak intensities may be affected by different extinction corrections in the different polarization channels while the phonon is not. 
It is also worthwhile to note that the temperature dependence of $2R_z$ is not to be taken as a quantity proportional to the magnetization since we are dealing with a non-Bravais lattice and the individual $\langle S^z_d \rangle$ are weighted by $e^{i\mathbf{Q}\cdot\mathbf{d}}$ and the corresponding sum is dependent on $\textbf{Q}$.
Above $T_{Curie}$, $2R_z$ is not zero due to the fact that the magnetic field is applied in the paramagnetic state and therefore, the values of $\langle S^z_d \rangle$ are finite. 
Above 260 K, the background was measured for each temperature at $\bf{Q}$=(-0.3, 0, 2) in both $\sigma^{++}_{z}$ and $\sigma^{--}_{z}$ channels, where they are found to be equal as expected.
This allows one to estimate the nuclear, $N$, and magnetic, $M_z$, (here magnetovibrational) scattering amplitudes ; the temperature population factor is also taken into account as a correction.
The obtained data are shown in Fig.3b.
The nuclear part is almost constant consistently with the weak variation of the Debye-Waller factor in a small temperature range with respect to our limited statistics.
The magnetic part rises as ferromagnetic ordering is increasing.
From this plot the origin of the phonon intensity cancellation at low temperature is clear : the magnetic amplitude increases when temperature decreases and accidentally reaches a value very close to the nuclear scattering amplitude. 
This causes the cancellation of the cross section $\sigma^{- -}_z$.

\begin{figure}
\centering
\vspace{-1cm}
\includegraphics[width=20cm]{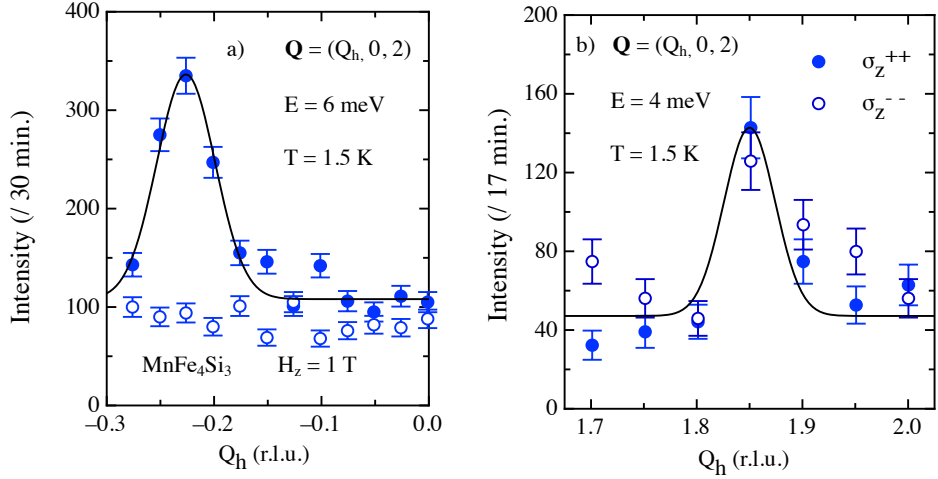}
\vspace{-8.5cm}
\caption{a) Inelastic neutron scattering spectra obtained in the $\sigma^{++}_z$ and $\sigma^{--}_z$ polarization channels at 1.5 K for a) $\textbf{Q}$=($Q_h$, 0, 2) for an energy transfer of 6 meV  b) $\textbf{Q}$=($Q_h$, 0, 2) for an energy transfer of 4 meV  with $\textbf{G}$=(2, 0, 2). Solid lines are Gaussian fits.}
\end{figure}

At low temperatures, it is found that the full TA branch shows such a behavior (See Fig.4 for data at $E$= 6 meV), except in the vicinity of the zone boundary.
This can be understood in the range of $\textbf{q}$ (i) where the acoustic approximation is valid and (ii) where the wave-vector dependence of the form factor can be neglected (see above).
In order to illustrate the fact that the phonon intensity cancellation is fortuitous due to an accidental coincidence of the nuclear and magnetic amplitudes, similar data were taken in another Brillouin zone for $\textbf{G}$=(2, 0, 2).
Figure 4b shows a corresponding INS spectra along $\bf{Q}$= ($Q_h$, 0, 2) at 4 meV and 1.5 K. The very limited statistics of the data impose to fix the peak position according to the one obtained in Fig.1. At this position, the phonon intensity is similar in the $\sigma_z^{++}$ and $\sigma_z^{--}$ channels.

Finally different branches were also investigated : the LA[100] phonon and TA[001] phonon polarized along $a^{*}$ around $\textbf{G}$=(3, 0, 0) : interference effects are observed without total cancellation of intensity.
Figure 5a shows INS spectra measured at $\bf{Q}$=($Q_h$, 0, 0) for $E$= 4 meV and $T$=1.5 K and illustrates the imbalance of intensities between the two polarization channels $\sigma^{++}_{z}$ and $\sigma^{--}_{z}$ for the corresponding LA[100] mode measured near $\textbf{G}$=(3, 0, 0). The panel b) shows the temperature dependence of the scattering amplitude for the LA[100] mode measured near $\textbf{G}$=(3, 0, 0). In contrast to what is shown in Fig.3b, the nuclear amplitude is much stronger than the magnetic one. Therefore, the interference effect is less pronounced and there is no total cancellation of intensity in $\sigma^{--}_{z}$ (Fig.5a).
The overall different behaviors of the different measured modes can be semi-quantitatively rationalized by comparing nuclear and magnetic structure factors for each Brillouin zone center $\textbf{G}$ (established for $H$=0 T \cite{Hering}), their ratio being of about 28 for $\textbf{G}$=(3, 0, 0), 6 for $\textbf{G}$=(2, 0, 2), 4 for $\textbf{G}$=(1, 0, 2) and 1 for $\textbf{G}$=(0, 0, 2).
\begin{figure}[h]
\centering
\vspace{-1cm}
\includegraphics[width=20cm]{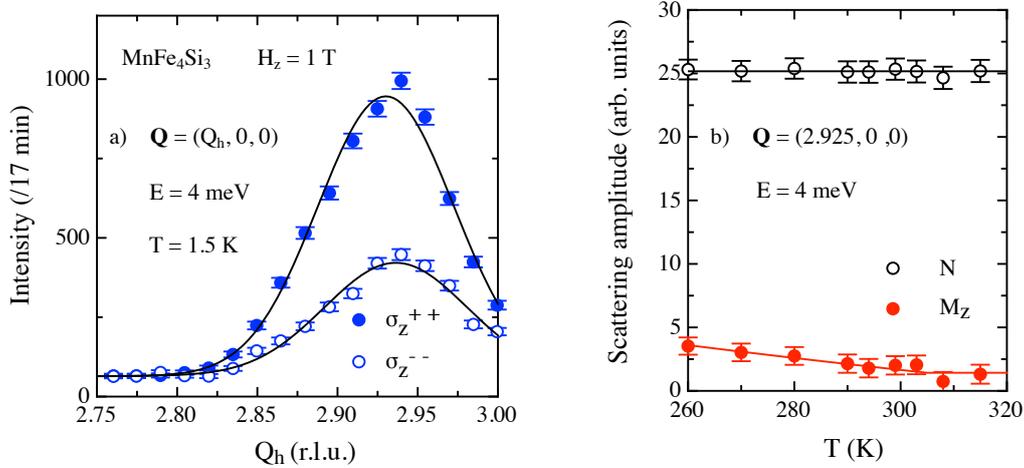}
\vspace{-8.5cm}
\caption{a) Inelastic neutron scattering spectra obtained in the $\sigma^{++}_{z}$ and $\sigma^{--}_{z}$ polarization channels at 1.5 K for $\textbf{Q}$=($Q_h$, 0, 0) for an energy transfer of 4 meV. Solid lines are Gaussian fits.  b) Temperature dependence of the nuclear and magnetovibrational scattering amplitudes obtained from the phonon intensity at $\bf{Q}$= (2.925, 0, 0) at $E$ = 4 meV. Lines are guides for the eyes.}
\end{figure}
\section{Conclusions}

Magnetovibrational scattering is often grasped as a source of spurious scattering \cite{Fernandez}. In very rare cases, it has been used as a probe of magnetism or as a probe of spin-lattice coupling.
As already mentioned, magnetic form factors as a function of $\bf{q}$ were obtained from polarized INS phonon measurements in Fe and Ni  \cite{Steinsvoll} and in the Fe$_{65}$Ni$_{35}$  invar alloy \cite{Brown1}.
Another interesting case also concerning Fe$_{65}$Ni$_{35}$ illustrates how new features occur when the approximation leading to the magnetovibrational scattering breaks down. Namely a forbidden phonon mode was explained by the modulation of the magnetic moment associated with compressive strain \cite{Brown2}. In the present paper, we report a total cancellation of phonon intensity for a given mode in a given Brillouin zone in longitudinal polarized INS measurements. While surprising at first glance, the effect arises purely from interferences between neutron scattering amplitudes and does not involve any specific physics. 
Put in a broad context, this large difference of cross-sections between $\sigma^{++}_{z}$ and $\sigma^{--}_{z}$ could be quite misleading especially for the INS experiments carried out with only one spin flipper where only either $\sigma^{++}_{z}$  or $\sigma^{--}_{z}$ is measured (together with a corresponding spin-flip cross-section).
The ignorance of such effect could lead to erroneous conclusions.
If one has in mind the archetypal  cancellation of the static nuclear and magnetic structure factors that makes an Heusler alloy a good neutron polarizer, crudely speaking, the same effect is realized here for a phonon, except that the magnetic scattering is replaced by the magnetovibrational scattering.
Conversely,  the next step would be to use the sensitivity of the NMI evidenced here to detect new physics associated with spin-lattice coupling beyond the magnetovibrational approximation as highlighted for Fe$_{65}$Ni$_{35}$ \cite{Brown2}.
To this respect magnetocaloric compounds could constitute an interesting playground for this search.
Last but not least, it is to be mentioned that NMI terms that are inelastic in the magnetic system are still to be discovered and several theoretical suggestions were made for the occurrence of such terms in complex magnetic systems \cite{Maleyev}.

\section{Acknowledgments}
We thank L.-P. Regnault for enlightening comments.

\section{Data availability}
All relevant data are available from the corresponding authors.\\ 
INS data collected at the ILL are available at https://doi.ill.fr/10.5291/ILL-DATA.CRG-2444.

\end{document}